# Exploring Cross-Domain Data Dependencies for Smart Homes to Improve Energy Efficiency

Shamaila Iram
University of Huddersfield
Huddersfield, United Kingdom
S.Iram@hud.ac.uk

Terrence Fernando
University of Salford
Salford, United Kingdom
T.Fernando@salford.ac.uk

May Bassanino
University of Salford
Salford, United Kingdom
M.N.Bassanino@salford.ac.uk

## ABSTRACT

Over the past decade, the idea of smart homes has been conceived as a potential solution to counter energy crises or to at least mitigate its intensive destructive consequences in the residential building sector. Smart homes have emerged as one of the applications of Internet of Things (IoT) that enabled the use of technology to automate and customize home services with reference to users' preferences. However, the concept of smart homes is still not fully matured due to the weak handling of diverse datasets that can be exploited to promote more adaptive, personalised, and context aware capabilities. Furthermore, instead of just deploying integrated automated services in the homes, the focus should be to bring the concerns of potential stakeholders into consideration. In this paper, we have exploited the concepts of ontologies to capture all sorts of data (classes and their subclasses) that belong to one domain based on stakeholders' requirements analysis. We have also explored their significant associations with other datasets from another domain. In addition, this research work provides an insight about what sorts of interdependencies exist between different datasets across different ontological models such as Smart homes ontology model and ICT ontology model.

## 1 INTRODUCTION

Smart homes have gained a tremendous interest from the research community, homes owners, energy providers, housing agencies as well as from the government agencies in the past few years. The rapidly increasing energy consumption rate poses an alarming threat to the worldwide environmental sustainability and economic stability. International Energy Agency's (IEA) statistics reveal that 32% of the total final energy is being consumed by the buildings [1]. This percentage is even higher in non-industrial areas. In recent years, energy efficiency and saving strategies have become a priority objective for energy policies due to the proliferation of energy consumption and $CO_2$ emission in the built environment. According to statistics [2], globally, 40% of all primary energy is being consumed in and by the buildings and contributes 30% of the $CO_2$ emissions [3]. In many developed countries, the buildings are considered as highest energy consumption sector than transportation and industry. For instance, in 2004, building sector has consumed 40%, 39%, and 37% of total primary energy in USA, the UK and the European Union [2]. The fact that how people consumes energy depends on human behaviour and other social, economic, environmental and geographical factors.

Data plays a significant role in the manipulation and exploitation of smart homes applications and services. Unfortunately, tremendous amount of data that is being generated in smart homes domain is usually not properly collected, structured and stored under their relevant domains. This eventually raise more issues in seamless integration, exchange and reusability of data for its adaptation and manipulation to multiple services [4]. This research work discusses the potential benefits of formally modelling the data for smart home services to deal with rich metadata and their semantics. This research work also contributes to the existing work by conceptualising the idea of extending modelling paradigm from smart home data to modelling the tools and techniques that belong to one particular concept within ICT domain. Determining the interrelationship between two domains i.e., smart home and ICT, and realizing the interdependencies among multiple concepts inside their domain could potentially facilitate seamless exchange of data and information to deploy significant services.

In order to address the issues relating to data handling, data exploitation and data visualization in smart homes applications, this research work is conducted with the intension of seeking answers to the following questions:

**Table 1: Functional demands of different stakeholders in designing and developing the Energy Efficient Smart Buildings**

| Stakeholders Analysis | | | |
|---|---|---|---|
| Occupiers/householders | Energy Providers | Housing Agencies/ Landlords | Government/policy makers |
| 1. Increased comfort level<br>2. Tracking energy consumption<br>3. Energy consumption patterns per day/month/year<br>4. Load shifting<br>5. Cost estimation<br>6. Cost Effectiveness<br>7. Customer Behaviour-Impact analysis<br>8. Anomaly Detection | 1. Demand and supply monitoring<br>2. Demand and supply balance<br>3. Load prediction<br>4. Transformer load management -Load balancing<br>5. Energy consumption patterns per day/month/year<br>6. Environmental impact - Eco impact<br>7. Decarbonisation | 1. Increased comfort level<br>2. Relative humidity rate- to avoid condensation, Dump or Mould<br>3. Cost estimation<br>4. Cost effectiveness<br>5. Customer behaviour- Impact analysis<br>6. Energy consumption patterns<br>7. Demand and supply monitoring<br>8. Environmental Impact- Eco Impact | 1. Comfort level<br>2. Revenue protection<br>3. Saving energy<br>4. Energy usage per month or per year<br>5. Environmental Impact-Eco Impact<br>6. Decarbonisation |

What are the key interests of potential stakeholders in developing smart homes? What are the techniques to classify their requirements under their relevant domain concepts?

How can we capture, manipulate and exploit most appropriate data using formal data models such as ontologies? Apart from data, how can these ontological models help to classify the tools and technologies under their relevant concepts in ICT domains

What kind of graphical user interface is required to facilitate the interaction and filtration of smart homes cross-domain data?

## 2 STAKEHOLDER'S REQUIREMENT ANALYSIS

A continuous depletion in natural energy resources, their growing demand and their detrimental environmental impact due to gas emissions and global warming, manifest a dire need of reforming energy efficiency policies especially in building sectors. This also urges potential stakeholders such as occupants, energy providers, housing agencies, policy makers and environmental groups to set their goals that are in consistent with the government policies

As shown in Table 1, It is not surprising that most important factors that share the same importance across all stakeholders are householders' health/well-being and their comfort [3], [5], [6], [7]. This implies that humans are the central and integral part of whole phenomena and their comfort should be given highest priority while designing a new building or starting a retrofit process. Amasyali and El-Gohary in 2016 [8] conducted an online survey of 618 residential and office buildings in three different US states to capture occupants' perspective on seven energy related values/factors. They analysed that health/well-being was given highest priority followed by comfort level (thermal, visual and indoor air quality) together with other attributes such as environmental protection and energy cost saving. Moreover, tracking energy consumption patterns on regular basis helps to save energy, especially, in residential buildings [9]. Porse *et al.,* [10] carried out energy analysis of aggregated account-level utility billing data for energy consumption across over two million properties in Los Angeles. They claimed that tracking energy consumption patterns using account level energy usage data can help local governments devise conservation strategies.

Research [11, 12] shows the importance of estimating cost of consumed energy as well as saved energy together with cost effective energy saving strategies to achieve higher energy efficiency. Studies [13] also reveal that consumers even with an adequate knowledge of how to save energy and also with a professed desire to do that could still be unsuccessful to achieve this goal. Frederiks *et al.,* [13] believe that this could be linked with people's values, attitudes and intensions and their observable behaviours which is commonly agreed as *knowledge-action gap* and *value-action gap*. This is why analysing human behaviour impact is one of the key measures that concern some stakeholders such as occupiers and housing agencies to achieve energy efficiency. Stakeholders also consider load shifting strategies [14] and anomaly detection in energy consumption patterns significant factors to improve energy efficiency.

There is a growing concern about global warming and climate change which is considered an alarming threat to the ecosystem and to the people [15]. According to statistics, cities are responsible for 75-80% of climate change [16].

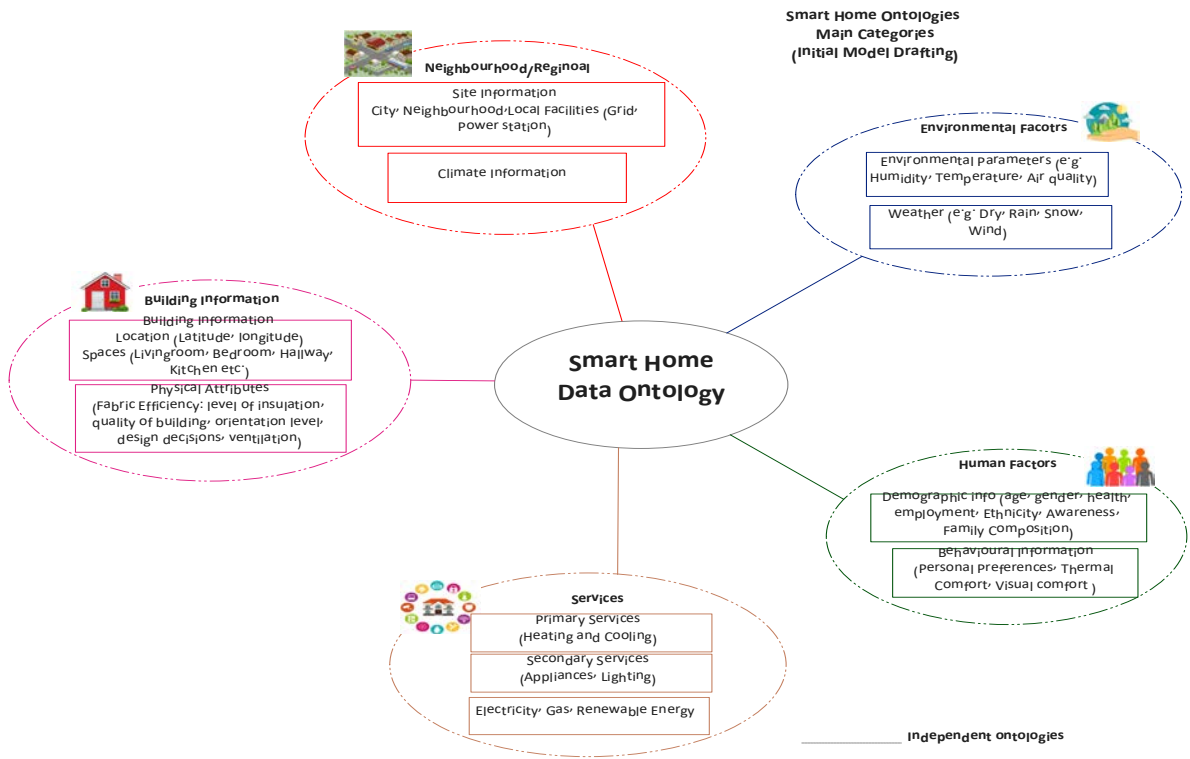

**Figure 1: Smart Home Data Ontology - Main Modules**

Therefore, environmental impact of the energy and also decarbonisation are considered important attributes by stakeholders that should be reflected in energy efficient smart buildings. Other variables that are of particular concern of energy provider are; demand side management [17] such as demand and supply monitoring and demand and supply balancing, transformer load prediction and load balancing. Similarly, some of the benefits that government is keen to achieve at strategic level from energy efficient buildings are; protecting revenues and saving energy as much as possible.

3 SMART HOMES DATA MODELLING- ONTOLOGY MODEL AND TAXONOMIES

Energy efficient system in smart homes needs an extensive knowledge base that accumulates large sets of data from heterogeneous sources to store all the required information. Smart home's data ecosystem constitutes all possible data concepts that can bring social and economic benefits to different stakeholders. A classical database model doesn't exhibit sufficient capabilities to capture significant details of a complex system domain [19]. Therefore, semantic representation of a particular domain with explicit detail of the main concepts and their inter-relationship is considered a priority to automate home services. Such practices enhance data processing capabilities such as data integration, data interoperability, and data analytical support. *Ontologies* are considered as one of the promising solutions to formally present data in a particular domain with a formal semantic language [19]. Ontology is an explicit, formal and shared specification of multiple concepts in a particular domain [20] defined with commonly agreed data structures, i.e., domain concepts, their attributes and the relations between them [4]. Web Ontology Language (OWL) which is one major technology of Semantic Web is used for the formal representation of knowledge, mainly, because of its formal definition and reasoning capabilities [30]. A *concept or class* in the ontology specifies the number of possible items required in a particular domain [18]. Fig.1 demonstrates all possible key concepts/classes that are required for the optimization of smart home energy system. Each single circle that is realized as a concept of smart home's domain itself presents an independent ontology on its own. The proposed ontology model for smart homes includes five domains such as Building Information, Neighbourhood/Regional Information, Environmental Factors, Human Factors, and Services.

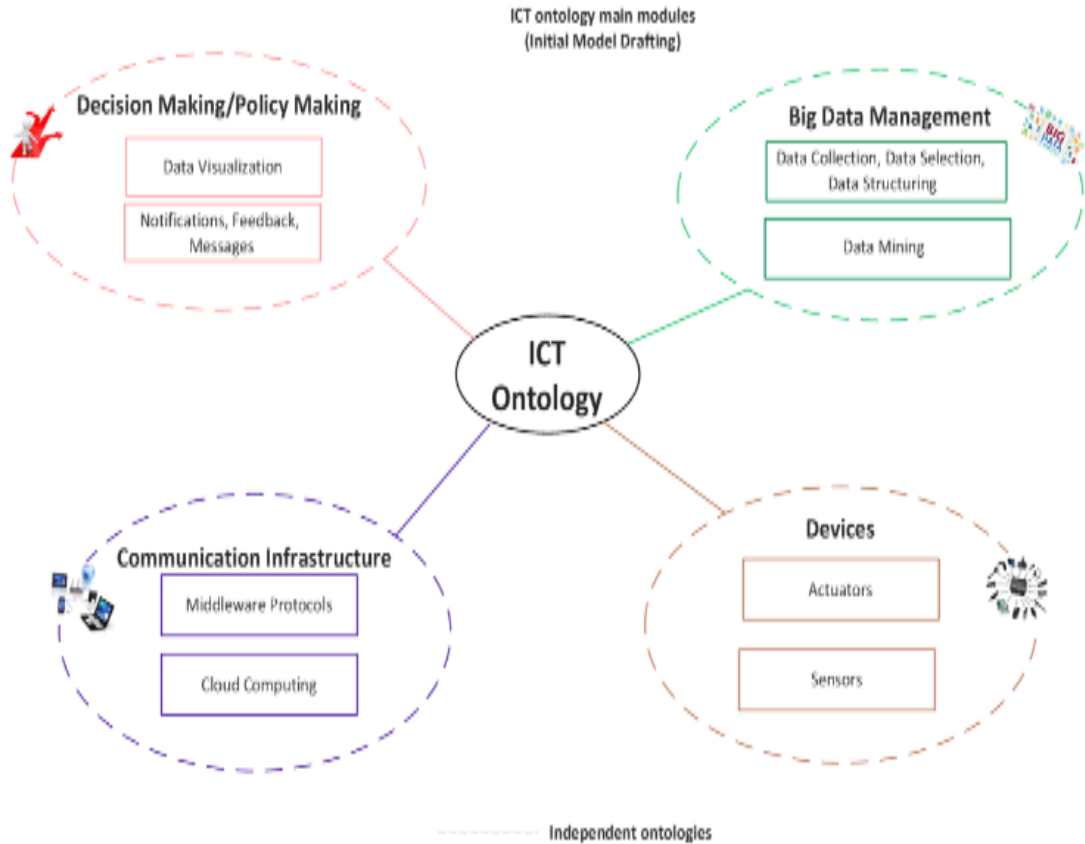

Figure 2: ICT Ontology - Main Modules

Each domain is further categorised into classes, sub classes and instances/features. For instance, *Neighbourhood/Regional* domain encapsulates two classes; *Site information* and *Climate information*. *Site Information* constitutes three subclasses; *City, Neighbourhood* and *Local facilities*. Furthermore, each subclass has instances/features such as Neighbourhood is further attributes as *Lower Layer Super Output (LSOA)* and *Local facilities* attributes are *Grid (Nearest substation)* and *Sub stations*.

*Building information* constitutes four classes such as *Address*, *Building Spaces*, *Building Resources*, and *Basic Information*. *Address* can be recognized as *Latitude and Longitude* values. Similarly, *Building Space* has four sub classes; *Livingroom, Bedroom, Hallway,* and *Kitchen*. *Building Resources* contains three sub classes; *Lighting, Heating* and *Appliances*. *Heating* is further divided to *Heating System* and *Distribution System*. *Heating System* can be attributed as *Combo Boiler*, *System Boiler*, and *Back Boiler* while *Distribution System* contains features such as *No. of Radiators*, *Size of Radiators* and *Water Tank*. *Appliance* has *Application Mode* which is further divided to *Off*, *On* and *Stand By*. *Basic Information* constitutes seven sub classes such as *EPC Rating*, *Air Test*, *Archetype*, *Ownership*, *Building Age*, *BIM Model*, *and Physical Attributes*. *Physical Attributes* means *Fabric Efficiency* which can be measured with the attributes/features such as *Level of Insulation*, *Quality of Building*, *Orientation Level*, *Design Decisions* and *Ventilation*.

Similarly, *Services* constitutes four classes; *Primary service*, *Secondary service* and *Energy*. *Primary services* are; *Heating* and *Cooling*. *Secondary services* are; *Appliances* and *Lighting*. However, *Energy* can be measured from *Electricity*, *Gas* and *Renewable energy usage*.

Forth domain that has significant importance in the area of smart homes is Human. *Human factors* have two sub classes; *Demographic information* and *Behavioural information*. *Demographic information* is further attributed as; *Age*, *Gender*, *Occupation*, *Awareness*, *Health status*, *Ethnicity*, and *Family composition*. *Family composition* could further be categorized as *Single*, *Couple*, and *Couple with children*. *Couple with children* is branched out to *Children with primary age*, *Children with secondary age* and other possible scenarios. However, *Behaviour*

*information* depends on *Personal preferences*, *Thermal comfort* and *Visual comfort*. *Personal preferences* could be measured by people's *Attitude* which depends on their *Financial* and *Ethical attributes*.

Last potential domain in the area of smart homes could be *Environmental factors*. It has two sub classes namely, *Environmental parameter* and *Weather*. *Environmental parameter* depends on the measures such as Temperature, Humidity, Air quality and Noise level. *Air quality* could be measured by *Pollution level* and *Pollen level*. *Pollution* further can be attributed as; *Carbon mono oxide* (CO), *Nitrogen di oxide* (NO), *Volatile organic components* (VOC) and *Particulates*. *Particulate*s may be the *Dust* or *Smoke particles*.

*Weather conditions* could further be recognized as *Dry*, *Rain*, *Snow*, and *Wind*. *Rain* can be measured by *Rainfall*, Snow by *Snowfall* and *Wind* by *Wind speed* or *Direction of speed over time*. Similarly, ICT ontology has been created to capture the key concept of technology domain as shown in the Fig.2. The ontology model of ICT domain constitutes five distinctive domains such as Big Data Management domain, Devices domain, Communication Infrastructure domain, and Decision Making/ Policy Making Domain.

## 3.2 Interdependences between ontologies-A case Study

This section demonstrates the communication pattern between smart homes ontology model as well as ICT ontology to implement a case study in the context of smart homes. For instance, users in a home might not be able to keep track of all the equipment at homes that are stayed turned on or on a stand-by mode overnight. However, by keeping knowledge of the usual behaviour of the users, an automated system can help to achieve energy optimisation in the home. For example, if there are least chances for a user to come back to the room, the unnecessary devices can be completely switched off to save energy. In order to implement this case study, the occupancy sensors from the ICT domain will check the presence of a person in the room.

Alternatively, this information could also be retrieved from the historical data sets by checking the normal behaviour of the people living in that house. If no occupancy is predicted in the room, then system will start reasoning mechanism using ontological models to shut down all unnecessary devices in the room. A detailed communication model between ICT ontology as well as homes ontology to implement the aforementioned use case is demonstrated in Fig.3 as:

1. Occupancy sensors will check the presence of a person in the room that contains one of the standby devices.
2. If no occupancy is detected in the room, system will check the historical data in the database to understand if an occupant is predicted in the room.
3. If no occupancy is predicted, the system will switch off the plugged in devices to save energy.

## 4 CONCLUSIONS AND FUTURE WORK

This research work has demonstrated the importance of capturing, managing and manipulating the right IoT data to improve energy efficiency in the context of smart homes. A secondary research has been conducted to capture and analyse stakeholders' requirements to understand the key themes that are relevant to each smart home's stakeholder. Based on stakeholders' requirements, authors proposed smart homes ontological model as well as ICT ontological model demonstrating key concepts in each domain explicitly. Furthermore, a communication model is presented to understand the data dependencies among different domains to implement a case study to automatically disconnect unnecessary devices from homes to save energy.

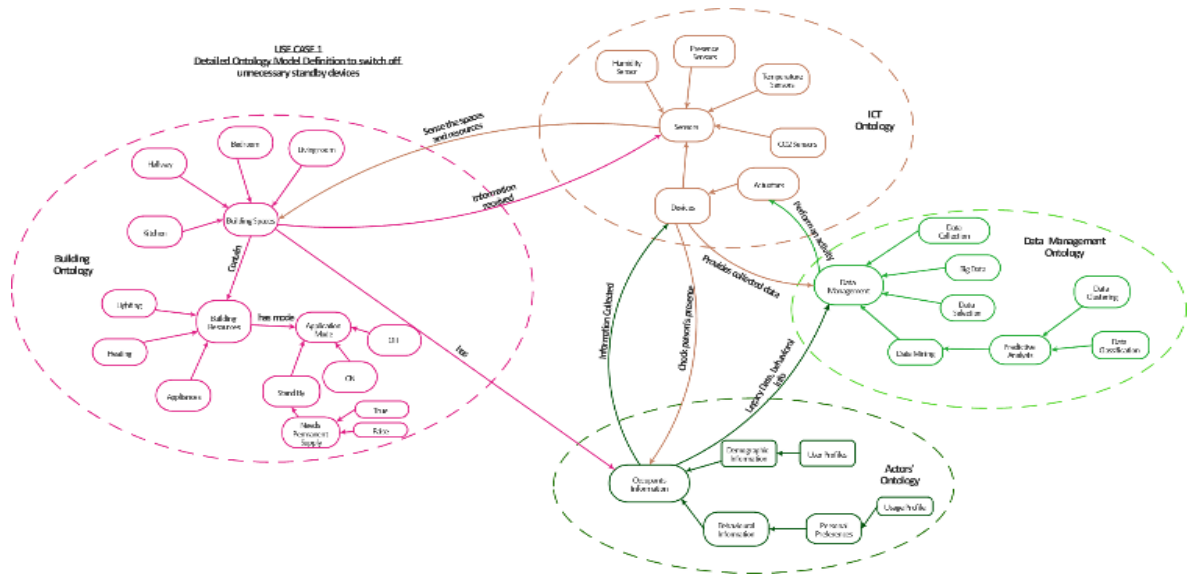

**Figure 3:** Communication model between Smart home ontology and ICT ontology

In this ongoing research work, future activities will involve the visualization of the ontological datasets into a graphical user interface to understand the data dependencies from different domains. Furthermore, the proposed system will be evaluated with focus on multiple smart homes applications.

## ACKNOWLEDGMENTS

Authors are very thankful to the Applied Buildings and Energy Research Group from University of Salford for their continuous feedback and support to improve this research work.